\documentclass[%
 reprint,footinbib,
superscriptaddress,
 floatfix,
 amsmath,amssymb,
 aps,
prapp, 
longbibliography,
]{revtex4-2}
\usepackage{graphicx}
\usepackage{physics}
\usepackage{amsmath}
\usepackage{natbib}
\usepackage{soul}
\usepackage{filecontents}
\usepackage{color}
\usepackage{dcolumn}
\usepackage[colorlinks=true, pdfstartview=FitV, linkcolor=blue, citecolor=blue, urlcolor=blue]{hyperref} 
\usepackage{bm}
\usepackage{hyperref}

\begin{document}

\preprint{APS/123-QED}

\title{Spin detection via parametric frequency conversion in a membrane resonator}

\author{Jan Ko\v{s}ata}
\email{kosataj@phys.ethz.ch}
\affiliation{Institute for Theoretical Physics, ETH Z{\"u}rich, 8093 Z{\"u}rich, Switzerland}
\author{Oded Zilberberg}
\affiliation{Institute for Theoretical Physics, ETH Z{\"u}rich, 8093 Z{\"u}rich, Switzerland}
\author{Christian L. Degen}
\affiliation{Laboratory for Solid State Physics, ETH Z{\"u}rich, 8093 Z{\"u}rich, Switzerland}
\author{R. Chitra}
\affiliation{Institute for Theoretical Physics, ETH Z{\"u}rich, 8093 Z{\"u}rich, Switzerland}
\author{Alexander Eichler}
\affiliation{Laboratory for Solid State Physics, ETH Z{\"u}rich, 8093 Z{\"u}rich, Switzerland}

\date{\today}

\begin{abstract}
Recent demonstrations of ultracoherent nanomechanical resonators introduce the prospect of developing protocols for solid state sensing applications. Here, we propose to use two coupled ultracoherent resonator modes on a Si$_3$N$_4$ membrane for the detection of small nuclear spin ensembles. To this end, we employ parametric frequency conversion between nondegenerate modes. The nondegenerate modes  result from coupled degenerate resonators, and the parametric conversion is mediated by periodic inversions of the nuclear spins in the presence of a magnetic scanning tip. We analyze potential noise sources and derive the achievable signal-to-noise ratio with typical experimental parameter values. Our proposal reconciles the geometric constraints of optomechanical systems with the requirements of scanning force microscopy and brings forth a promising platform for spin-phonon interaction and spin imaging.

\end{abstract}
\maketitle

\section{\label{sec:level1}Introduction}
Nanoscale Magnetic Resonance Imaging (NanoMRI) aims at detecting nuclear spins in three spatial dimensions with sub-nanometer resolution~\cite{Sidles_1991, Degen_2009, Poggio_2010, Mamin_2013, Staudacher_2013, Loretz_2014}. In contrast to other techniques like electron microscopy or X-ray tomography, NanoMRI is able to obtain 3D images of complex macromolecules in a nondestructive manner. Combined with the elemental selectivity of MRI, this emerging technique has the potential to become a unique probe of the 3D composition of nanostructures.

Achieving the necessary sensitivity to detect the magnetic moment of a nanometer-sized nuclear spin ensemble is a formidable task. One candidate technique to achieve this goal is Magnetic Resonance Force Microscopy (MRFM)~\cite{Sidles_1991, Rugar_2004, Degen_2009, Nichol_2012, Nichol_2013, Moores_2015, Rose_2018}. In MRFM, nuclear spins are periodically inverted inside a magnetic field gradient to generate a force proportional to the spin magnetic moment. A mechanical transducer is used to detect this force and to translate it into an optical or electrical signal. The sensitivity of the transducer is typically limited by the thermomechanical force noise power spectral density (PSD) of its resonant mode,
\begin{align} \label{eq:1}
    S_f = 4 k_B T \gamma,
\end{align}
where $k_B$ is the Boltzmann constant and $T$ is the mode temperature. The transducer's dissipation coefficient is $\gamma = \sqrt{m k}/Q = m \omega_0/Q$, where $m$ is the effective mass of the mode, $\omega_0=2\pi f_0$ is the angular resonance frequency, $k = m\omega_0^2$ is the spring constant, and $Q$ is the mechanical quality factor. In order to reach better sensitivity, much effort is being invested to reduce the dissipation~\cite{Mamin_2001, Moser_2013, Tao_2014, Weber_2016, Reinhardt_2016, Norte_2016, Rossi_2017, Delepinay_2017, Tsaturyan_2017, Heritier_2018, deBonis_2018, Ghadimi_2018}.

Traditional MRFM setups are constructed around cantilever resonators with very small spring constants~\cite{Mamin_2001, Moores_2015, Heritier_2018, Nichol_2012, Nichol_2013, Rose_2018, Rossi_2017}. While this strategy reduces $\gamma$, it can generate issues with long-term stability and strong tip-sample interaction. In addition, one of the primary goals of MRFM is the imaging of biological samples and macromolecules, which are difficult to mount on the tip of a cantilever. Recently, a route towards ultra-low damping coefficients has emerged through the development of soft-clamped silicon nitride membranes and strings, with localized defect modes that feature quality factors up to the $Q \sim 10^9$ range ~\cite{Tsaturyan_2017, Ghadimi_2018, Rossi_2018}. Thanks to this outstanding virtue, silicon nitride resonators offer force sensitivities comparable to those of singly-clamped cantilevers, in spite of their higher masses and resonance frequencies.

\begin{figure}
        \includegraphics[width=\columnwidth]{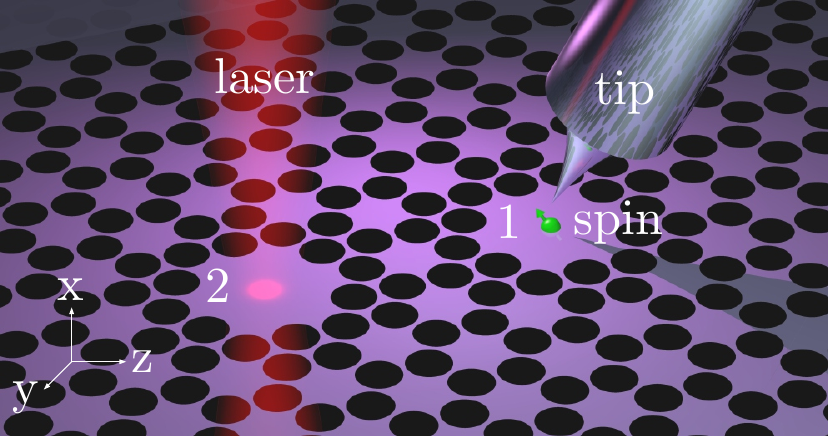} 
        \caption{\label{fig:fig1} Proposed measurement setup. A silicon nitride membrane is patterned such that a phononic bandstructure is obtained with explicit band gaps. Two unpatterned `defect' areas define in-gap vibrational modes (labeled 1 and 2). A sample spin (or an ensemble of spins) on defect $1$ is periodically inverted by radio-frequency pulses, while a sharp tip provides a magnetic field gradient, transducing the spin inversion to a force acting on the membrane. A cavity laser focused on defect $2$ is used for driving and readout. Each hole in the pattern has a diameter of roughly $80~\mu$m.}
        \end{figure}

Silicon nitride membranes are attractive transducers for spin detection instruments~\cite{Scozzaro_2016, Blankenhorn_2017, Takahashi_2018, Fischer_2019}. Their large surface allows simple placement of samples, and their high spring constants ensure low displacement drift and bending even in the proximity of a scanning tip. However, there remains one serious obstacle, which we term the `frequency mismatch problem': the vibrational modes of interest are in the low MHz range. Inverting nuclear spins adiabatically at an angular rate of $2\omega_0$, as required by traditional MRFM protocols~\cite{Sidles_1991, Degen_2009}, is unrealistic with such devices. More specifically, current experiments use oscillating fields of a few mT in amplitude to achieve inversion rates of $\sim 10$~kHz~\cite{Grob_2019}. Scaling this method to MHz frequencies would entail oscillating field amplitudes that are incompatible with cryogenic operation and nanoscale precision.

In this paper, we propose a sensing scheme that circumvents the frequency mismatch problem. We employ a parametric upconversion method~\cite{Dougherty_1996, Moore_2010} to couple two nondegenerate normal modes. The coupling is achieved by a modulation of the effective mechanical frequencies via nuclear spins that are inverted at the splitting frequency. We explore how degenerate local defect modes on the membrane give rise to split extended normal modes, how parametric modulation of a local mode leads to coupling between the normal modes, and how parametric modulation is generated by nuclear spins placed on one local mode in the presence of a magnetic field gradient source. In addition to solving the frequency mismatch problem, the extended nature of the normal modes allows for spatial separation of the sample placement and the readout of the membrane vibrations. This facilitates the integration of our proposed scanning force setup into a high-finesse optical cavity, which allows very sensitive readout of the membrane vibrations and enables a host of optomechanical control techniques~\cite{Aspelmeyer_2014}. From our analysis with realistic experimental parameters, we currently predict a sensitivity competitive with that of contemporary cantilever-based MRFM~\cite{Moores_2015, Rose_2018, deWit2019, Grob_2019} while harnessing the advantages of the membrane platform (these we discuss in more detail in Section~\ref{subsec:disc}). Furthermore, we pinpoint the critical properties of the resonator to design transducers with  improved spin detection performances in the future.

The working principle of our sensing scheme is outlined in Section~\ref{sec:general}. In Section~\ref{sec:model}, we derive the full equation of motion for the system in the coupled mode basis. In Section~\ref{sec:linear}, we obtain a simple closed form expression for the signal gain in the absence of nonlinearities and noise. The limitations of this scheme due to the onset of nonlinear behavior are explored in Section~\ref{sec:nl}, followed by noise analysis and a derived expression for the signal-to-noise ratio in Section~\ref{sec:noise}. Finally, in Section~\ref{sec:disc} we present a survey of current state-of-the-art experimental possibilities and evaluate the expected performance of our method. 

\section{General idea} \label{sec:general}
	We now present our spin detection scheme based on membrane transducers. Consider an elastic membrane patterned with a hexagonal array of holes (see Fig.~\ref{fig:fig1}) to create a phononic band gap~\cite{Tsaturyan_2017}. Small defects in the pattern define localized out-of-plane vibrational modes whose frequencies lie within the gap - these modes are effectively isolated from the rest of the membrane and can thus reach extremely high quality factors. 
	
\begin{figure} [ht!] 
    \centering
    \includegraphics[width=\columnwidth]{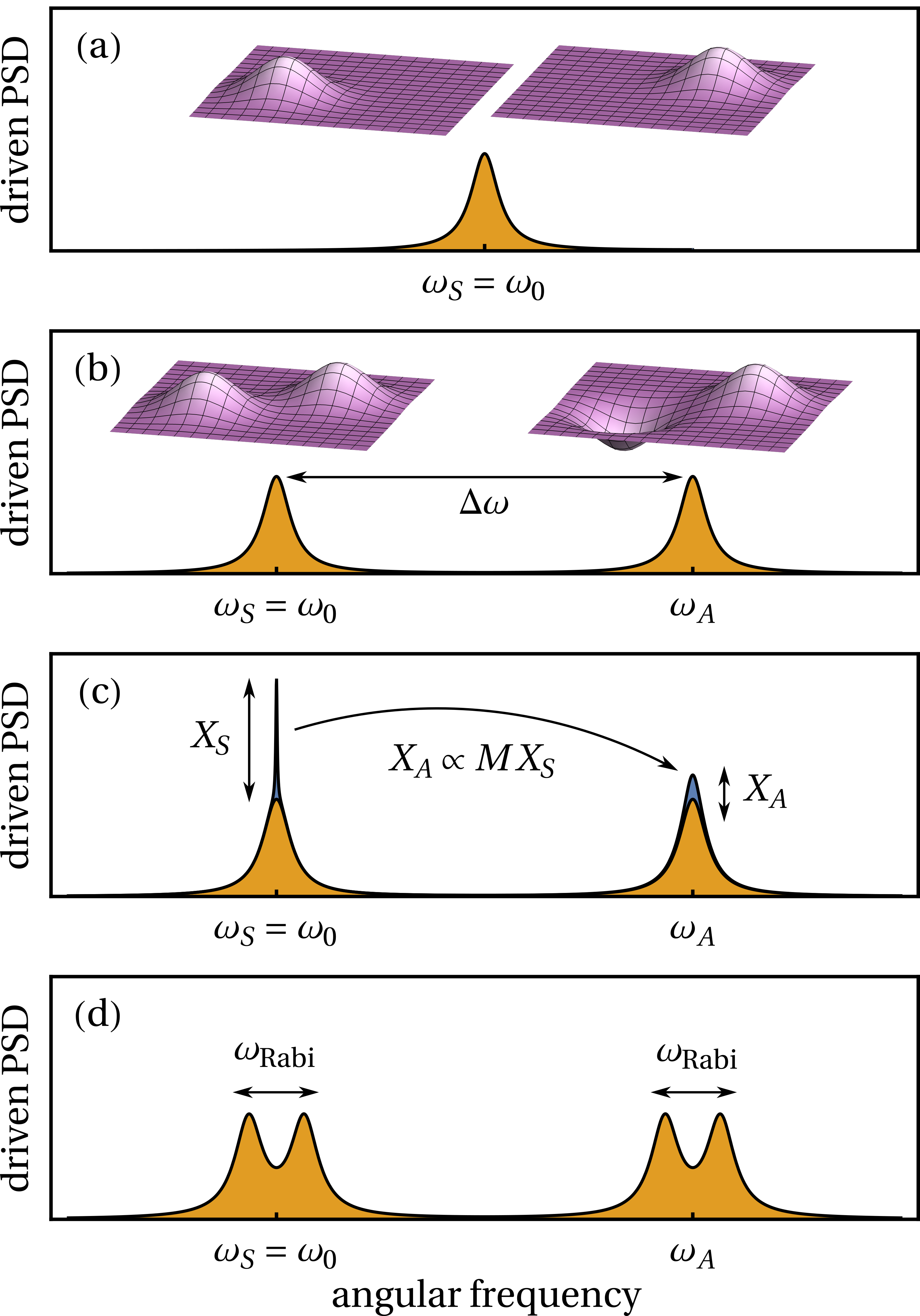}
    \caption{Mode power spectra at different levels of coupling. (a) The two uncoupled degenerate modes at $\omega_S$. Inserts provide schematic representations of the mode shapes. (b) frequency splitting introduced by linear spatial coupling. (c) spin-mediated amplitude at $\omega_A$ generated by weak parametric modulation and a strong drive at $\omega_S$. (d) further splitting (Rabi oscillations) due to strong parametric driving (not realized in our case).}
    \label{fig:splittings}
\end{figure}

	We consider two such modes with equal frequencies and effective masses. When in close proximity, the modes are mechanically coupled, giving rise to symmetric and antisymmetric normal modes with frequencies $\omega_S = \omega_0$ and $\omega_A > \omega_S$ [see Fig.~\ref{fig:splittings} (a) and (b)]~\cite{Catalini_2020}. The frequencies $\omega_S$ and $\omega_A$ are in the MHz range, but their difference $\Delta \omega = \omega_A - \omega_S$ is on the order of a few kHz. 
	
An ensemble of spins with magnetization $M$ is placed on one of the defects and is periodically inverted by radio-frequency pulses~\cite{Degen_2009}. In the presence of a magnetic tip, the magnetic field gradient couples mechanical motion to the spin moment. Namely, the second derivative leads to a frequency modulation of the corresponding defect mode. This modulation translates into a time-dependent coupling between the normal modes~\cite{Frimmer_2014}. If this coupling is periodically varied exactly at the rate $\Delta \omega$, it generates what is known as parametric frequency conversion or parametric mode coupling. When one of the modes, e.g. $\omega_S$, is additionally resonantly driven by an external force to amplitude $X_S$, the parametric mode coupling induces the antisymmetric mode at $\omega_A$ to be driven by the combination of $X_S$ and $M$ [cf. Fig.~\ref{fig:splittings} (c)] to amplitude $X_A$. The presence of the spins can thus be inferred from the oscillations, which can be read off at $\omega_A$ at either of the two defect locations. In this way, a slow spin inversion can lead to a detectable signal at a high-frequency mode.
 
\section{Model description} \label{sec:model}

We model the system as two equivalent interacting resonators with coordinates $x_1$ and $x_2$, corresponding to modes 1 and 2 in Fig.~\ref{fig:fig1}. Note that $x_1$ and $x_2$ are projections of the full motion of the 2D system onto the normal mode basis~\cite{Catalini_2020}. Our treatment is entirely classical, as in our case thermal effects overcome the oscillator's energy level spacing, i.e.,  $\hbar\omega_0 \ll k_B T$ down to cryogenic temperatures (with $\hbar$ the reduced Planck constant). Mechanical nonlinearities are added, as they turn out to play an important role (cf. Sec.~\ref{sec:nl}). We designate $x_1$ to host a time-dependent magnetic moment $\vb{M}(t)$ in the presence of an inhomogeneous magnetic field $\vb{B}(x_1)$ in the $x$ direction. The goal of our scheme is to measure the magnetization $\abs{\vb{M}(t)}$, which corresponds to the net polarization of a nanoscale ensemble of nuclear spins. Additionally, $x_2$ is externally driven by a force $F(t)$, e.g., by laser radiation pressure. The corresponding Hamiltonian reads
\begin{equation}
H = H_0 + H_{1},
\end{equation}
where $H_0$ describes two non-interacting nonlinear resonators, and
\begin{equation} \label{eq:hamiltonian}
H_{1} = \frac{m \omega_0 \Delta \omega}{2} (x_1 - x_2)^2  - \vb{M}(t) \vdot \vb{B}(x_1) - F(t)\: x_2
\end{equation}

contains the linear coupling between the resonators, the field interaction, and the external drive, respectively. The component of spin polarization that is perpendicular to the external field undergoes Larmor precession and does not contribute to the signal. Therefore $\vb{M}(t) \vdot \vb{B}(x_1) = M(t) B(x_1)$ and the corresponding equations of motion in the presence of homogeneous dissipation $\gamma$ are
\begin{multline} \label{eq:x1}
\ddot{x}_1 + \gamma \dot{x}_1 + \omega_0^2\: [1 + \chi_1(t)] \:x_1 + \omega_0 \Delta \omega (x_1 - x_2) +K(x_1)\\
= \frac{1}{m} \bigg[M(t)\frac{\partial}{\partial x_1} B(x_1) +  \xi_1(t)\bigg], 
\end{multline}
and
\begin{multline} \label{eq:x2}
\ddot{x}_2 + \gamma \dot{x}_2 + \omega_0^2\: [1 + \chi_2(t)] \:x_2 +  \omega_0 \Delta \omega (x_2 - x_1) +K(x_2)\\
= \frac{1}{m} [F(t) + \xi_2(t)].
\end{multline}
The function $K(x)$ contains all nonlinear elements, cf. Sec.~\ref{sec:nl}. The terms $\xi_{1,2}(t)$ and $\chi_{1,2}(t)$ represent thermal and frequency noise, respectively, whose roles are discussed in Sec.~\ref{sec:noise}.

The magnetization $M(t)$ fluctuates with a finite correlation time, known as the spin relaxation time $T_1$ in NMR experiments. Simultaneously, we apply periodic spin-flipping with frequency $\Omega$, so that
\begin{align} \label{eq:M(t)}
M(t) = M \xi_M(t) \cos{\Omega t} \\\expval{\xi_M(t) \xi_M(t')} = e^{- \abs{t-t'} / T_1}
\end{align}
where $\xi_M(t)$ is a stochastic term with an approximately Lorentzian PSD in the frequency domain. In accord with the frequency mismatch problem discussed above, the spin-flipping rate is slow relative to the membrane mode frequency, $\Omega \ll \omega_0$.

Taking the equilibrium point to be $x_1 = 0$ and expanding $B(x_1)$ yields, to second order \footnote{Note that higher order terms introduce nonlinearities into the equation of motion, but here these are negligible in magnitude.},
\begin{equation} \label{eq:diff_B}
\frac{\partial  B(x_1)}{\partial x_1} = \eval{\frac{\partial  B(x_1)}{\partial x_1}}_{x_1 = 0} + \eval{\frac{\partial^2 B(x_1)}{\partial x_1^2}}_{x_1 = 0} x_1.
\end{equation}

Substituting Eq.~\eqref{eq:diff_B} into Eq.~\eqref{eq:x1}, the first derivative of $B$ corresponds to a direct drive of mode 1 by the spin force, which is measured in conventional MRFM by flipping the spin at the resonator frequency, $\Omega = \omega_0$. For $\Omega \ll \omega_0$, this drive is off-resonant and can be neglected. The second term corresponds to a force that is proportional to the amplitude $x_1$, i.e., it corresponds to a parametric drive~\cite{Rugar_1991, Lifshitz_2008, Leuch_2016}. This parametric drive is also strongly detuned from the main parametric resonance frequency, which is $\Omega = 2 \omega_0$. The frequency shift caused by this term has been observed in experiments~\cite{Rugar_2004, Garner_2004, Degen_2009, Lee_2012}, but its practical use for nuclear spin detection is typically hindered by frequency noise in the resonator. 
\\

The key role of the higher field derivatives in Eq.~\eqref{eq:diff_B} is brought out in the normal mode basis (see Fig.~\ref{fig:splittings} for a visualization). Applying the linear transformation
\begin{equation} \label{eq:transform}
\mqty(x_S \\ x_A) = \frac{1}{\sqrt{2}}\mqty(1 && 1 \\ 1 && -1) \mqty(x_1 \\ x_2)
\end{equation}
yields two normal modes labeled $x_S$ (symmetric) and $x_A$ (antisymmetric), governed by the two equations of motion
\begin{multline} \label{eq:bigeom}
\ddot{x}_i + \gamma \dot{x}_i +\omega_i^2(t) \: x_i + K_i(x_i, x_j) = \\ \delta\omega^2(t) x_j + \frac{F(t)}{\sqrt{2} m} + \frac{\xi_i(t)}{m},
\end{multline}
with $i,j \in \{S,A\}, \: i \neq j$. The normal modes are split in frequency~\footnote{Please note that the elastic coupling energy term proportional to $(x_1-x_2)^2$ does not affect the symmetric mode while it raises the frequency of the antisymmetric mode. The more common situation where the coupling is proportional to $x_1 x_2$ (e.g.,~Rabi splitting) results in a symmetric splitting around $\omega_0$.},
\begin{align} \label{eq:normal_basis}
\begin{split}
\omega_S^2(t) &= \omega_0^2 - \delta\omega^2(t), \\
\omega_A^2(t) &= \omega_0^2 + 2 \omega_0 \Delta \omega- \delta\omega^2(t),
\end{split}
\end{align}
where we denoted
\begin{equation} \label{eq:deltaomega}
\delta\omega^2(t) = \frac{M(t)}{2 m} \partial^2_x B,
\end{equation}
and from now on we use the short-hand notation $\partial^2_x B \equiv \eval{\frac{\partial^2 B(x_1)}{\partial x_1^2}}_{x_1 = 0}$. 

The transformed nonlinear terms $K_i(x_i, x_j)$ further couple the two modes. Note that for resonators with non-identical frequencies and/or masses, a transformation analogous to Eq.~\eqref{eq:transform} can always be found which cancels the linear coupling term.
 
\section{Linear case} \label{sec:linear}
Before turning to the key performance characteristic of our proposed scheme - the signal-to-noise ratio (SNR) - in Sec.~\ref{sec:noise}, we demonstrate its working principle on a simple case. To this end, we neglect nonlinearities and noise terms in Eq.~\eqref{eq:bigeom} and set $T_1 \rightarrow \infty$.

It is important to note that $\delta\omega^2(t)$ enters as two different effects. First, it enters Eq.~\eqref{eq:normal_basis} as a (parametric) modulation in time of the normal mode frequencies. As mentioned before, this modulation is far detuned from resonance and can be neglected. Second, $\delta\omega^2(t)$ appears in Eq.~\eqref{eq:bigeom} as an explicit coupling term between $x_S$ and $x_A$~\cite{Frimmer_2014}.

Energy upconversion is equivalent to a driving force exerted by one mode onto the other. Taking, without loss of generality, $x_S$ as the strongly-driven `pump' and $x_A$ as the `readout' mode, we set the external force to be $F(t) = F \cos(\omega_S t)$ and write the pump mode amplitude as
\begin{equation} \label{slowflow}
x_{S}(t) = X_S \cos{\omega_{S} t}.
\end{equation}
The oscillating term $\delta\omega^2(t)x_S(t)$ in Eq.~\eqref{eq:bigeom} now acts as a driving force for $x_A$ that facilitates a frequency conversion: the low-frequency parametric drive $\delta\omega^2(t)$ of the resonator $x_1$ is upconverted into two driving terms acting on the mode $x_A$ at frequencies $\omega_S \pm \Omega$. Flipping the spins at the modes' frequency difference, $\Omega = \Delta \omega$, hence results in a resonant driving force for $x_A$ with amplitude
\begin{equation} \label{eq:fspin}
F_{\rm spin} = \frac{M \partial_x^2 B}{4} X_S.
\end{equation}
In this way, the driven pump mode generates a magnetization-dependent motion of the readout mode. We can thus detect $M$ by measuring the Fourier component $\hat{x}_A(\omega_A)$ (denoted $X_A$), which is proportional to $F_{\rm spin}$ and has a fixed ratio to the pump mode amplitude $X_S$,
\begin{equation} \label{eq:moderatio}
X_A = \frac{M \partial^2_x B}{4 m \omega_A^2} Q \,X_S.
\end{equation}
where $Q = \omega_0 / \gamma$.

Note that strong parametric coupling of two nondegenerate modes generates doubly-split states [see Fig.~\ref{fig:splittings} (d)]~\cite{Okamoto_2013, Abdo_2013}. This can be understood as a periodic redistribution of energy between the modes, akin to Rabi oscillations~\cite{Frimmer_2014, Faust_2013, Okamoto_2013_NP}, which manifest as a beating in the amplitude of each normal mode. The frequency of the Rabi oscillations for a sinusoidal parametric drive $\delta\omega^2(t) \equiv \delta\omega^2\sin\left(t\Delta\omega\right)$ is given by the corresponding natural frequency shift, $\omega_{Rabi} = \delta \omega^2 / 2\omega_0$. In our case, however, $\omega_{Rabi} \ll \gamma \ll \omega_0$, meaning that energy upconverted from $\omega_S$ to $\omega_A$ is dissipated long before it can be coherently transported back to $\omega_S$. We thus neglect coherent Rabi oscillations and only look for steady-state amplitudes of the normal modes $x_S$ and $x_A$ under the influence of weak energy upconversion [Fig.~\ref{fig:splittings} (c)].

In summary, we can see that the strong drive at frequency $\omega_S$ has been converted into a signal at $\omega_A$ that depends linearly on $M$, i.e., it corresponds to the (instantaneous) magnetization of the measured spin or spin ensemble. Measuring the ratio in Eq.~\eqref{eq:moderatio}, instead of a directly-driven resonator amplitude, enables the use of high-frequency resonators as MRFM sensors. Note that for finite spin lifetimes this magnetization will turn into a fluctuating quantity whose variance in time represents the signal~\cite{Grob_2019}.

\section{Nonlinear effects} \label{sec:nl}
Inspecting Eq.~\eqref{eq:moderatio} suggests that, in order to maximize the signal amplitude, we should drive the pump mode as much as possible. However, nonlinearities, which are naturally present in any real system, become non-negligible as the oscillation amplitudes increase. Though such coupled nonlinear equations are difficult to solve, we will see that the effect of nonlinearities on our sensing scheme can be quantified in a straightforward manner.

In line with the contemporary development of defect modes in membrane nanoresonators~\cite{Catalini_2020}, we consider a combination of a Duffing term $\alpha$ and nonlinear damping terms $\Gamma_1$ and $\Gamma_2$, such that $K$ in Eqs.~\eqref{eq:x1} and~\eqref{eq:x2} becomes
\begin{equation}
K(x_i) = \frac{\alpha}{m} x_i^3 + (\Gamma_1 x_i^2 + \Gamma_2 \dot{x}_i^2)\: \dot{x}_i, \quad i = 1,2,
\end{equation}
with our analysis being directly applicable to higher order nonlinear terms. 

 Identifying again $x_S$ as the pump mode, we refer to the equations of motion [Eq.~\eqref{eq:bigeom}] in the limit of $x_S \gg x_A$. The nonlinearity of $x_S$ will shift its resonant frequency and induce motion at higher harmonics of the driving frequency~\cite{Rand_2005}. These higher harmonics generally affect the readout mode $x_A$ non-resonantly, such that the approximation $x_S \approx X_S \cos(\omega_S t)$ remains correct. The readout mode $x_A$ has a much smaller amplitude than $x_S$ because it is not driven resonantly. We can therefore continue to treat it as a linear resonator,
\begin{multline} \label{xasimp}
\ddot{x}_A + 
\gamma^{\rm {nl}}_A(t)\:  \dot{x}_A 
+ \omega^{\rm {nl}}_A (t)^2 x_A \\
= X_S \: \delta \omega^2(t) \cos(\omega_S \: t)+ \frac{F(t)}{\sqrt{2} m}.
\end{multline}
However, nonlinearities in the bare resonators $x_1, x_2$ couple the motions of $x_S$ and $x_A$ (cf. Appendix~\ref{app:nonlin}). The damping term $\gamma^{\rm {nl}}_A$ and natural frequency $\omega^{\rm {nl}}_A$ are hence affected by the large amplitude $X_S$ as 
\begin{align} \label{eq:omeganl}
\begin{split}
\gamma^{\rm {nl}}_A(t) &= \gamma + \frac{X_S^2}{4} \big[\Gamma_1 + 3 \omega_S^2 \Gamma_2 +(\Gamma_1 - 3 \omega_S^2 \Gamma_2) \cos{2 \omega_S t}\big],\\
\omega^{\rm {nl}}_A(t)^2 &= \omega_A^2[1+\lambda_1 (1 + \cos{2 \omega_S t}) + \lambda_2\sin{2 \omega_S t}],
\end{split}
\end{align}
where
\begin{equation} \label{eq:lambdas}
\lambda_1 = \frac{3 \alpha}{4 m \omega_A^2} X_S^2,\quad \text{ and} \quad \lambda_2 = \frac{\Gamma_1 \omega_S}{2 \omega_A^2} X_S^2 .
\end{equation}
The time-dependent terms in Eq.~\eqref{eq:omeganl} act as off-resonant parametric drives and have no significant effect (cf. Appendix~\ref{app:parametric}). Similarly, the natural frequency shift introduced by the Duffing nonlinearity $\alpha$ is negligible ($\lambda_1 \cong 10^{-6}$), precluding significant changes in the response of $x_A$~\cite{Lifshitz_2003}. Note that $\lambda_1$ converts any noise present in the amplitude $X_S$ to frequency noise of $x_A$~\cite{Kenig_2012}. Importantly, though, the effective damping is increased. Defining a nonlinear damping parameter $\Gamma_{\rm nl} = \frac{1}{4} (\Gamma_1 + 3 \omega_S^2 \Gamma_2)$, the quality factor of the readout mode $x_A$ is lowered to
\begin{equation} \label{Qnl}
Q_{\rm nl} = \frac{\omega_0}{\gamma + \Gamma_{\rm nl} X_S^2},
\end{equation}
which in turn limits the signal gain, cf. Eq.~\eqref{eq:moderatio} with $Q\rightarrow{Q}_{\rm nl}$. As we will show in Sec.~\ref{sec:noise}, the increased damping is also detrimental to the SNR since it increases the thermal noise power, in accordance with the fluctuation-dissipation theorem, [cf. Eq.~\eqref{eq:1}].

\section{Statistical treatment} \label{sec:noise}
\subsection{Noise terms in the model}
In a realistic setting, the bare resonators [Eqs.~\eqref{eq:x1},~\eqref{eq:x2}] are subject to both additive thermal white noise $\xi_{1,2}(t)$ and multiplicative frequency noise $\chi_{1,2}(t)$~\cite{Cleland_2002}. In the frequency domain $\xi_{1,2}$ is spectrally flat, whereas $\chi_{1,2}$ will usually drop off as $\omega^{-1}$ or $\omega^{-2}$~\cite{Fong_2012}. The observed amplitude of the readout mode $x_A$ is thus a combination of the desired signal and fluctuations in the system. To obtain the SNR, we analyze the additive and multiplicative noise components in Eq.~\eqref{eq:bigeom} separately.

\textbf{Transformed noise terms.} In the normal mode basis, the multiplicative noise is simply
\begin{equation}
\chi_{A,S} = \frac{\chi_1(t) + \chi_2(t)}{2},
\end{equation}
while the additive noise has contributions from both additive (thermal) and multiplicative (frequency) noise of the bare oscillators,
\begin{equation}
\xi_{A,S} = \frac{\xi_1(t)\pm\xi_2(t)}{\sqrt{2}}+ \frac{m\omega_0^2[\chi_2(t) - \chi_1(t)] }{2} x_{S,A}. \\
\end{equation}
Specifically, we note the coherent term $x_S$ entering the additive noise $\xi_A$ - taking $x_S$ to be the monochromatic pump [Eq.~\eqref{slowflow}], this upconverts the noise from $\chi_1(t)$ and $\chi_2(t)$, creating a qualitatively different noise term in the coupled system.

\textbf{Noise PSD.} For equivalent bare oscillators, we assume equal frequency noise PSDs, $S_{\chi_1}(\omega) = S_{\chi_2}(\omega) \equiv S_\chi(\omega)$, whereas the thermal noise PSD is taken constant as per the equipartition theorem, $S_{\xi_1} = S_{\xi_2} \equiv S_\xi$. We then arrive at the PSD of $\xi_A$,
\begin{equation} \label{eq:SxiA}
S_{\xi_A}(\omega) = \frac{2\omega_0 m k_B T}{\pi Q_{\rm nl}} + \frac{(m \omega_0^2 X_S)^2}{8} [S_{\chi}(\omega - \omega_S) + S_\chi(\omega + \omega_S)],
\end{equation}
 comprising both thermal noise and upconverted frequency noise \footnote{Note that as we are using a single-sided PSD as a function of angular frequency $\omega$, the thermal noise appears with a factor of $2/\pi$ instead of the more familiar case of a PSD as a function of frequency $f = \omega/2\pi$, which results in a factor $4$}. The last term in Eq.~\eqref{eq:SxiA} will typically be negligible due to the fast decay of $S_\chi(\omega)$ with $\omega$.

\textbf{Effect of frequency noise.} In general, colored frequency noise is difficult to treat analytically. Exact results have been obtained for dichotomous and trichotomous Markovian noise with a Lorentzian PSD~\cite{Bourret_1973, Gitterman_2003, Mankin_2008}. These display a highly complex dependence of the system response on the noise profile, which however only manifests at relatively long coherence times. In our case, the typical noise coherence times $\tau_c$ fall within the limit $\gamma \ll \tau_c^{-1} \ll \omega_0$. It has been shown~\cite{Gitterman_2003} that in this case, the noise simply shifts the system slightly off resonance, so that the response [Eq.~\eqref{eq:moderatio}] decreases to
\begin{equation} \label{eq:varomega}
X_A \rightarrow X_A \bigg[1 - \frac{\text{Var}(\omega_A^2)}{\omega_0^4}Q^2\bigg].
\end{equation}
The frequency variance $\text{Var}(\omega_A^2)$ can be obtained via the Wiener-Khinchin theorem by integrating the power spectral density of the dimensionless term $\chi(t)$,
\begin{equation} 
\text{Var}(\omega_A^2) = \omega_0^4 \int_0^\infty S_\chi(\omega) \:d\omega.
\end{equation}

Apart from the intrinsic frequency noise $\chi$, our system is also affected by the conversion of any noise in the pump amplitude $X_S$ into the frequency noise of $x_A$, as shown in Appendix~\ref{app:nonlin}. However, both sources result in negligible corrections compared to the effect of thermal noise acting on $x_A$ and we disregard them from now on.

\textbf{Effect of spin fluctuations.} We now study the impact of the finite spin lifetime. During the course of a measurement, the magnetization will fluctuate with a characteristic time $T_1$, cf. Eq.~\eqref{eq:M(t)}. In the frequency domain, this corresponds to a Lorentzian distribution. The signal (i.e., the force PSD due to the nuclear spins) is therefore broadened to give
\begin{equation}\label{eq:S_spin}
S_{\rm spin}(\omega) = F^2_{\rm spin} \frac{T_1}{\pi[1 + (\omega-\omega_A)^2 T_1^2]},
\end{equation}
where $F_{\rm spin}$ is the force originating from parametric driving with a coherent magnetic moment [Eq.~\eqref{eq:fspin}]. From a practical perspective (cf. Sec.~\ref{sec:disc}), it is desirable to increase the bandwidth of the resonator beyond the spin lifetime, $2 Q / \omega_A < T_1$~\cite{Degen_2007}. This is routinely achieved by active feedback damping~\cite{Courty_2001, Kleckner_2006, Poggio_2007, Rossi_2018}.

\textbf{General displacement PSD.} We finally present a formulation of the displacement PSD of the readout mode in the presence of feedback damping and various noise sources. Since we consider the case $k_B T  \gg \hbar\omega_A$, we neglect zero-point fluctuations and the discrete nature of the energy spectrum. 

We start by defining the susceptibility of the mode as
\begin{equation} \label{eq:susc_damped}
g^2(\omega) = \frac{1 / m^2}{(\omega^2 - \omega_A^2)^2 + (\omega \omega_A / Q_{\rm fd})^2}
\end{equation}
with $Q_{\rm fd} = Q_{\rm nl}/(1+p)$ being the (nonlinear) quality factor damped by a feedback gain $p$.
The mode is driven by the fluctuating force $S_{\xi_A}(\omega)$ defined in Eq.~\eqref{eq:SxiA} and is further subject to detector noise $S_{\rm det}$ and to quantum backaction force noise $S_{\rm qba}$, which represents the non-negligible disturbance of the system by an increasingly precise measurement~\cite{Courty_2001, Clerk_2010}. The latter takes on the value 
\begin{equation} \label{eq:S_ba}
S_{\rm qba} = \frac{\hbar^2}{4 \pi^2 S_{\rm det} \eta}
\end{equation}
with $0 < \eta \leq 1$ being the detection efficiency~\cite{Rossi_2018}. The observed displacement PSD of the readout mode in the presence of all of these fluctuating forces as well as a spin signal becomes~\cite{Poggio_2007, Rossi_2018}
\begin{equation} \label{eq:S_x}
\begin{split}
S_x(\omega) = g^2(\omega)[& S_{\xi_A}(\omega) + S_{\rm qba} \\& + S_{\rm spin}(\omega) + g_{p=0}^{-2}(\omega) S_{\rm det}]
\end{split}
\end{equation}
where $g_{p=0}^{2}(\omega)$ is the susceptibility without feedback damping. There are three important points to note here: first, we can see from Eq.~\eqref{eq:S_x} that feedback damping decreases the thermomechanical displacement noise PSD but not the underlying force noise PSD (terms in the bracket on the right-hand side). The benefit of feedback damping for nuclear spin detection is only to allow for rapid sampling of statistically independent spin configurations~\cite{Degen_2007}. Second, the fact that zero-point fluctuations are reduced by feedback damping does not violate the Heisenberg uncertainty principle, since the added measurement uncertainty corresponds to at least one half quantum of energy~\cite{Courty_2001, Rossi_2018}. Third, tuning $S_{\rm det}$, for instance by varying the laser power in an optical cavity used to detect the resonator motion, enables an optimal measurement that minimizes $S_x$ over a desired bandwidth~\cite{Clerk_2010}.

\textbf{Filtering.} In order to reduce the measured displacement noise, we apply a filter to reject noise far off the signal frequency. As a concrete example, we consider a standard Butterworth filter, which modulates the signal with $G(\omega) = [1 + ((\omega - \omega_A)/\omega_f)^{2 n}]^{-1/2}$, where $\omega_f$ denotes the bandwidth and $n$ the filter order. The signal of our experiment corresponds to the displacement variance driven by the spin signal,
\begin{equation} \label{eq:signal}
\sigma_{\rm spin}^2 = \int_{0}^{\infty} G(\omega)^2  g(\omega)^2 S_{\rm spin} (\omega) \: d\omega,
\end{equation}
which is measured together with a noise background of
\begin{multline} \label{eq:noise}
\sigma_{\rm noise}^2 = \int_{0}^{\infty} G(\omega)^2 g(\omega)^2[S_{\xi_A}(\omega) \\+ S_{\rm qba} + g_{p=0}^{-2}(\omega) S_{\rm det}]\: d\omega.
\end{multline}

\begin{figure*}
    \centering
    \includegraphics[width=180mm]{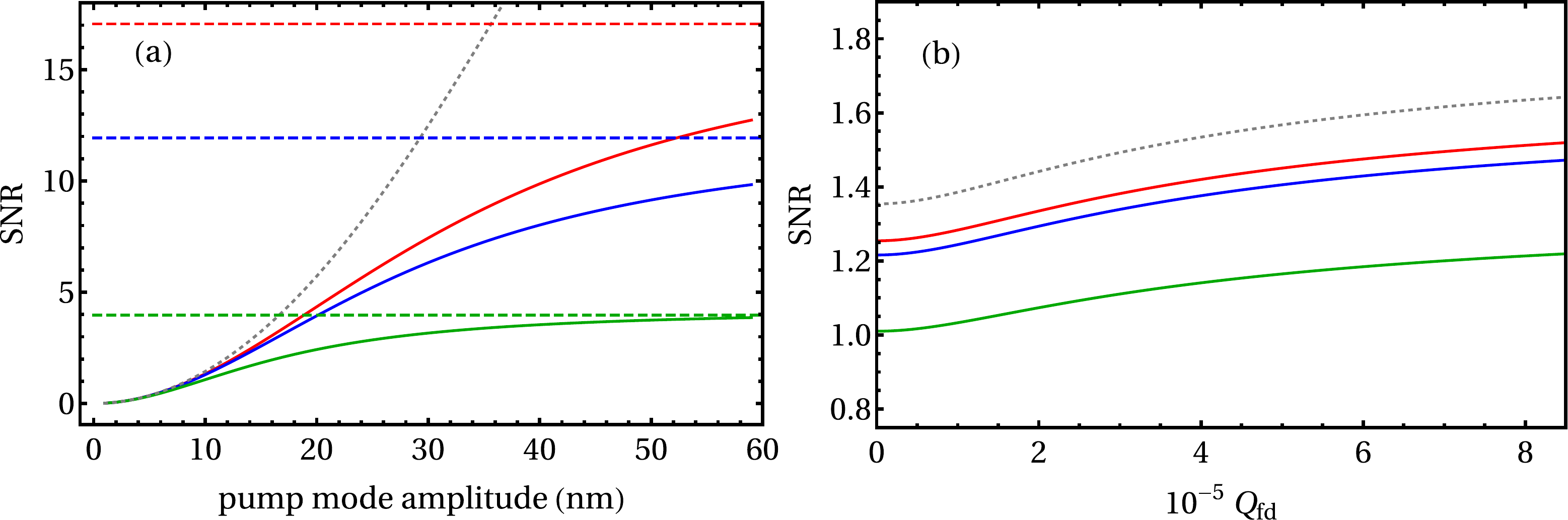}
    \caption{The SNR [cf. Eq.~\eqref{eq:SNR}] as a function of (a) pump mode amplitude $X_S$ and (b) feedback-damped quality factor $Q_{\rm fd}$. The lines correspond to different values of nonlinear damping $\Gamma_{\rm nl}$ (from top to bottom, red: $7\cross10^{13}$ m$^{-2}$s$^{-1}$, blue: $1\cross10^{14}$ m$^{-2}$s$^{-1}$, green: $3\cross10^{14}$ m$^{-2}$s$^{-1}$, dotted gray: no nonlinear damping). The collection time is $t_c = 240$~s and we consider an ensemble size of $10^4$ proton spins at $T = 0.2$ K, as well as $T_1$ = 50 ms, and $S_{\rm det} =  10^{-31}$~m$^2$\;s (cf. other values in Appendix~\ref{app:vals}). In (a), the quality factor is $Q_{\rm fd} =2 \cross 10^5$, and the dashed lines show the respective asymptotic limits [cf. Eq.~\eqref{eq:snrlimit}]. In (b), the pump mode amplitude is $10$~nm.}
    \label{fig:snr}
\end{figure*}

\section{Experiment proposal} \label{sec:disc}
We now assess the feasibility of the proposed measurement scheme using representative experimental parameters of patterned Si$_3$N$_4$ membranes, cf. Fig.~\ref{fig:fig1}. The full set of parameters is included in Appendix~\ref{app:vals}.

First of all, we need to relate $\sigma_{\rm spin}^2$ to the number of spins in the measured ensemble. To this end, we utilize the MRFM framework of spin variance sensing~\cite{Degen_2007}. For an ensemble of $N$ spins, as $N$ becomes small, the thermal (Boltzmann) polarization scales as $N$ and is eventually outweighed by the spin noise, whose standard deviation scales as $\sqrt{N}$. The preferred measurable quantity at the nanoscale is hence the magnetization variance
\begin{equation}
\expval{M^2 - \expval{M}^2} = N \mu^2,
\end{equation}
where $\mu = 1.4 \times 10^{-26}$~J\;T$^{-1}$ is the proton magnetic moment. Assuming $\expval{M} \cong 0$, the expected spin force is then $\expval{F^2_{\rm spin}} \propto N \mu^2$. Our aim is to estimate the variance by taking successive noisy readings of $M$ as it fluctuates in time, and hence determine $N$. For a spin ensemble with lifetime $T_1$ and matched filter bandwidth $\omega_f = 1 / T_1$, in the limits $\sigma_{\rm spin}^2 \ll \sigma_{\rm noise}^2$, we can represent the  SNR after a collection time $t_c \gg T_1$ in the concise form~\cite{Degen_2007}
\begin{equation} \label{eq:SNR}
\text{SNR} = \frac{1}{2} \sqrt{\frac{t_c}{T_1}}\frac{\sigma_{\rm spin}^2}{\sigma_{\rm noise}^2}.
\end{equation}

Eq.~\eqref{eq:SNR} is the most important characteristic of the proposed experiment.

\begin{figure*}
    \centering
    \includegraphics[width=180mm]{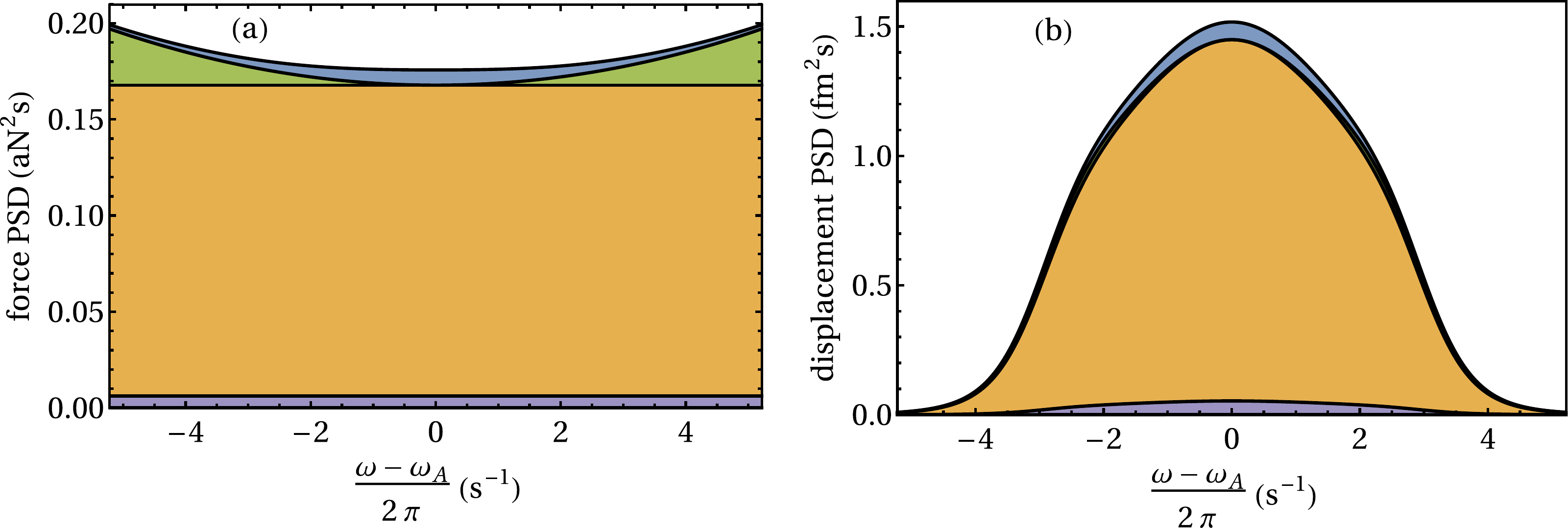}
    \caption{A cumulative plot of (a) the force PSD~\cite{Note1} acting on the readout mode [cf. Eq.~\eqref{eq:S_x}] and (b) the resulting filtered displacement PSD [cf. Eq.~\eqref{eq:noise}]. The individual contributions, from bottom to top, correspond to the quantum backaction [purple, cf. Eq.~\eqref{eq:S_ba}], effective thermal noise $S_{\xi_A}$ (orange), detector noise $S_{\rm det}$ (green), and the spin signal $S_{\rm spin}$ [blue, cf. Eq.~\eqref{eq:S_spin}]. We use conservative values for $S_{\rm det} = 10^{-31}$~m$^2$~s, and $T = 0.2$ K, $T_1$ = 50 ms, $Q_{\rm fd} =2 \cross 10^5$, $\Gamma_{\rm nl} = 1 \cross 10^{14}$ m$^{-2}$ s$^{-1}$, $X_S = 10$ nm. (Other values in Appendix~\ref{app:vals}.)}
    \label{fig:PSDs}
\end{figure*}

\subsection{Expected SNR}
Let us now evaluate the SNR of the feedback-damped system [Eq.~\eqref{eq:SNR}]. Driving the pump mode $X_S$ stronger boosts the SNR via the parametric conversion into the force $\abs{F_{\rm spin}}$ [cf. Eqs.~\eqref{eq:fspin} and~\eqref{eq:S_spin}]. However, it also increases the readout mode dissipation via nonlinear damping [Eq.~\eqref{Qnl}], which in turn increases thermal fluctuations [Eq.~\eqref{eq:SxiA}]. 

\textbf{Asymptotic limit.} The nonlinear damping eventually becomes the dominant dissipation mechanism, whereby for
\begin{equation} \label{eq:stronglim}
\Gamma_{\rm nl} \gg \gamma / X_S^2,
\end{equation}
both the signal and the thermal noise PSD scale as $X_S^2$, resulting in a limiting value of the SNR,
\begin{equation} \label{eq:snrlimit}
\lim_{X_S\rightarrow \infty}{\text{SNR}} =\frac{ C_n (\mu\: \partial_x^2 B)^2}{64 k_B T m \:\Gamma_{\rm nl} } N \sqrt{t_c \: T_1}.
\end{equation}
where
\begin{equation} \label{eq:cn}
C_n = \frac{\int_{0}^{\infty} [ (1 + z^{2n}) (1 + z^2)]^{-1} \: dz}{\int_{0}^{\infty} (1 + z^{2n})^{-1} \:dz}
\end{equation}
is a dimensionless constant, depending solely on the filter order $n$. Note that Eq.~\eqref{eq:snrlimit} is independent of the intrinsic linear damping parameter $\gamma$, assuming it is possible to drive the system strongly enough to satisfy the inequality in Eq.~\eqref{eq:stronglim}.

\textbf{Case study.} We proceed to calculate the expected SNR [Eq.~\eqref{eq:SNR}] for values motivated by recent experiments~\cite{Rossi_2018}. The filtering constant $C_n$ [Eq.~\eqref{eq:cn}] increases with $n$; we use the value $n=4$ ($C_4 \cong 0.77$) as higher orders bring negligible improvement. A plot of the SNR against the pump mode amplitude for three representative values of $\Gamma_{\rm nl}$ is shown in Fig.~\ref{fig:snr}. All parameters are taken from Appendix~\ref{app:vals} unless stated otherwise.

The current membrane devices typically possess $\Gamma_{\rm nl} \cong 1\cross10^{14}$ m$^{-2}$ s$^{-1}$~\cite{Catalini_2020}. In Fig.~\ref{fig:snr}, we see that this allows an SNR exceeding 1 at relatively modest pump mode amplitudes of $X_S = 10$ nm and collection times of 240 s, approaching 9.5 at stronger pumping. [Please note that the spatially resolved amplitudes of $x_{1,2}$ are equal to $X_S/\sqrt{2}$, cf. Eq.~\eqref{eq:transform}. For $X_S = 10$ nm, the spin sample moves therefore with an amplitude of $7.1$ nm.]. This projected performance is on par with current state-of-the-art MRFM experiments, although with a significant potential for improvement stemming from the unusual sensing mechanism (cf. Sec.~\ref{subsec:disc}).

A breakdown of the different noise sources in terms of their impact on the observed displacement is shown in Fig.~\ref{fig:PSDs}. The principal noise component is the thermomechanical noise.

\subsection{Discussion and conclusions} \label{subsec:disc}
The SNR results presented in Fig.~\ref{fig:snr} compare favorably to recent MRFM experiments. We expect to reach an SNR value of 1 after 240 seconds of measurement with ensemble sizes around $N \sim 10^4$ spins, which matches the sensitivity of current state-of-the-art measurements obtained with ultrasoft cantilevers and nanowires~\cite{Moores_2015, Rose_2018, Grob_2019}. Further significant improvements are expected since the design of ultracoherent nanoscale resonators is an area of active research. As shown in Eq.~\eqref{eq:snrlimit}, the instrumental limitation to the SNR depends on the product $(m \Gamma_{\rm nl})^{-1}$. On the one hand, low mass $m$ can readily be achieved by designing thin ribbons instead of a drum-mode resonator. On the other hand, the nonlinear damping coefficient $\Gamma_{\rm nl}$ has hitherto not received much attention. Optimizing the resonator design accordingly could open up unprecedented sensitivities in nuclear spin sensing.

Our scheme offers significant practical advantages over instruments based on cantilevers or nanowires. (i) The membrane surface is convenient for placement of `large' samples, such as viruses or biological molecules in the $100~$nm range. (ii) The spring constant of the membrane resonator modes is orders of magnitude higher than that of typical MRFM cantilevers, which results in a drastically reduced susceptibility towards spatially varying interaction potentials that affect the sensitivity. The scheme is easily extendable to oscillators with unequal masses and natural frequencies. (iii) Our parametric upconversion scheme does not require any electrical or magnetic signal at the frequency of the detection mode. This will be helpful to avoid spurious driving of the sensor which can make data interpretation difficult. (iv) Finally, the magnetic field gradient source in our scheme is located on the scanning tip, which will allow to utilize commercial magnetic force microscopy (MFM) probes. Membrane-based MRFM thus has the potential to become a mature and versatile NanoMRI platform.

In summary, we have theoretically demonstrated the feasibility of
using MHz optomechanical membrane resonators as force sensors for nuclear spins. Our work highlights the potential of membrane platforms for sensitive spin detection and should encourage further development of membrane-based NanoMRI instruments and spin-mechanics quantum information platforms.

\begin{acknowledgments}
For fruitful discussions and inspiration, we acknowledge A. Schliesser, Y. Tsaturyan, and L. Catalini. This work was supported by the Swiss National Science Foundation through grant CRSII5 $177198/1$, PP00P2$\_$163818.
\end{acknowledgments}

\newpage
\appendix

\section{The effect of nonlinearities on normal modes}
\label{app:nonlin}

The transformation in Eq.~\eqref{eq:transform} can be applied to the nonlinear equation of motion, although this no longer decouples $x_S$ and $x_A$. Taking the difference $x_1 - x_2$ from Eqs.~\eqref{eq:x1} and~\eqref{eq:x2}, we obtain, after rearranging and leaving out the noise terms,
\begin{multline} \label{eq:fullnonlin}
\ddot{x}_A + \bigg[\gamma + \frac{\Gamma_1}{2} (x_S^2 + x_A^2) + \frac{\Gamma_2}{2} (3 \dot{x}_S^2 + \dot{x}_A^2)\bigg] \dot{x}_A \\ 
+ \bigg[\omega_0^2 + 2 \omega_0 \Delta \omega + \frac{\alpha}{2 m} x_A^2 + \frac{3 \alpha}{2 m} x_S^2 + \Gamma_1 x_S \dot{x}_S \bigg] x_A  \\ 
= \frac{M(t) \partial_x^2 B}{2 m} (x_S + x_A) + \frac{F(t)}{\sqrt{2} m}\;.
\end{multline}

Since $x_S \gg x_A$, we can drop all higher-order $x_A$ terms in Eq.~\eqref{eq:fullnonlin}. The readout mode thus behaves as a linear oscillator, but with an $x_S$-dependent damping and spring constant. Nonlinearity of the mode $x_S$ itself plays no role in the detection mechanism, we can thus continue to write its amplitude as $x_S = X_S \cos(\omega_S t)$. Eq.~\eqref{eq:fullnonlin} hence simplifies to the form used in Sec.~\ref{sec:nl}.

We note that in general, the pump mode $x_S$ is subject to thermal noise. Looking at the prefactor of $x_A$ in Eq.~\eqref{eq:fullnonlin}, we see this is converted to frequency noise of $x_A$ via the Duffing nonlinearity $\alpha$~\cite{Kenig_2012, Villanueva_2013, Yurke_1995}. Let us describe the noisy pump by $X_S = (X_{S0} + \delta X_S)  \cos(\omega_S t)$, where the $\delta X_S$ is the stochastic contribution of thermal noise. To leading order, this affects the frequency of $x_A$,
\begin{equation}
    \omega_A^2 = \omega_0^2 + 2 \omega_0 \Delta \omega + \frac{3 \alpha}{m} X_{S0}  \, \delta X_S \cos^2(\omega_S t).
\end{equation}
Taking $\expval{\delta X_S} = 0$ and $\expval{\delta X_S^2} = k_B T /  m \omega_S^2$ and dropping the oscillatory off-resonant terms, this introduces a variance of the frequency $\omega_A^2$,
\begin{equation}
    \text{Var}(\omega_A^2) = \frac{3}{8} \left(\frac{3 \alpha}{m}\right)^2 \, X_{S0}^2 \, \expval{\delta X_S^2}.
\end{equation}
For the reference values in Appendix~\ref{app:vals} and $X_{S0} = 10$~nm, we obtain $\text{Var}(\omega_A^2) = 1.4\cross10^7$~s$^{-4}$. While this is far higher than the intrinsic frequency noise, it does not significantly diminish the resonant response of $x_A$ [cf. Eq.~\eqref{eq:varomega}].

\section{Spurious parametric terms in the nonlinear regime} \label{app:parametric}

Exciting the pump mode imparts multiple parametric drives on the readout mode, cf. Eqs.~\eqref{eq:omeganl} and~\eqref{eq:lambdas}. The effect of parametric driving is well-explored in the resonant case, where the spring constant is varied at twice the resonator's natural frequency~\cite{Cleland_2005}. The response amplitude in that case increases or decreases depending on the relative phase of the parametric and external drives, a phenomenon known as parametric squeezing~\cite{Lifshitz_2008}. 

In Eq.~\eqref{eq:omeganl} however, the spurious parametric terms oscillate at $2 \omega_S$ and are thus strongly detuned from $2\omega_A$. A straightforward perturbative treatment then shows that the drive induces spurious motion at frequencies $\abs{\omega_A \pm 2 \omega_S}$. The signal -- extracted from the Fourier component at $\omega_A$ -- is therefore unaffected. The result calculated earlier for the linear regime remains valid even in the presence of nonlinearities, with $Q$ replaced by $Q_{\rm nl}$. It is in principle possible for the drives to cause parametric instabilities, however, as our prospective system is far from the unstable regime, we do not pursue this issue further.

\section{Magnetic field simulations} \label{app:fields}
\begin{figure}
    \centering
    \includegraphics[width=\columnwidth]{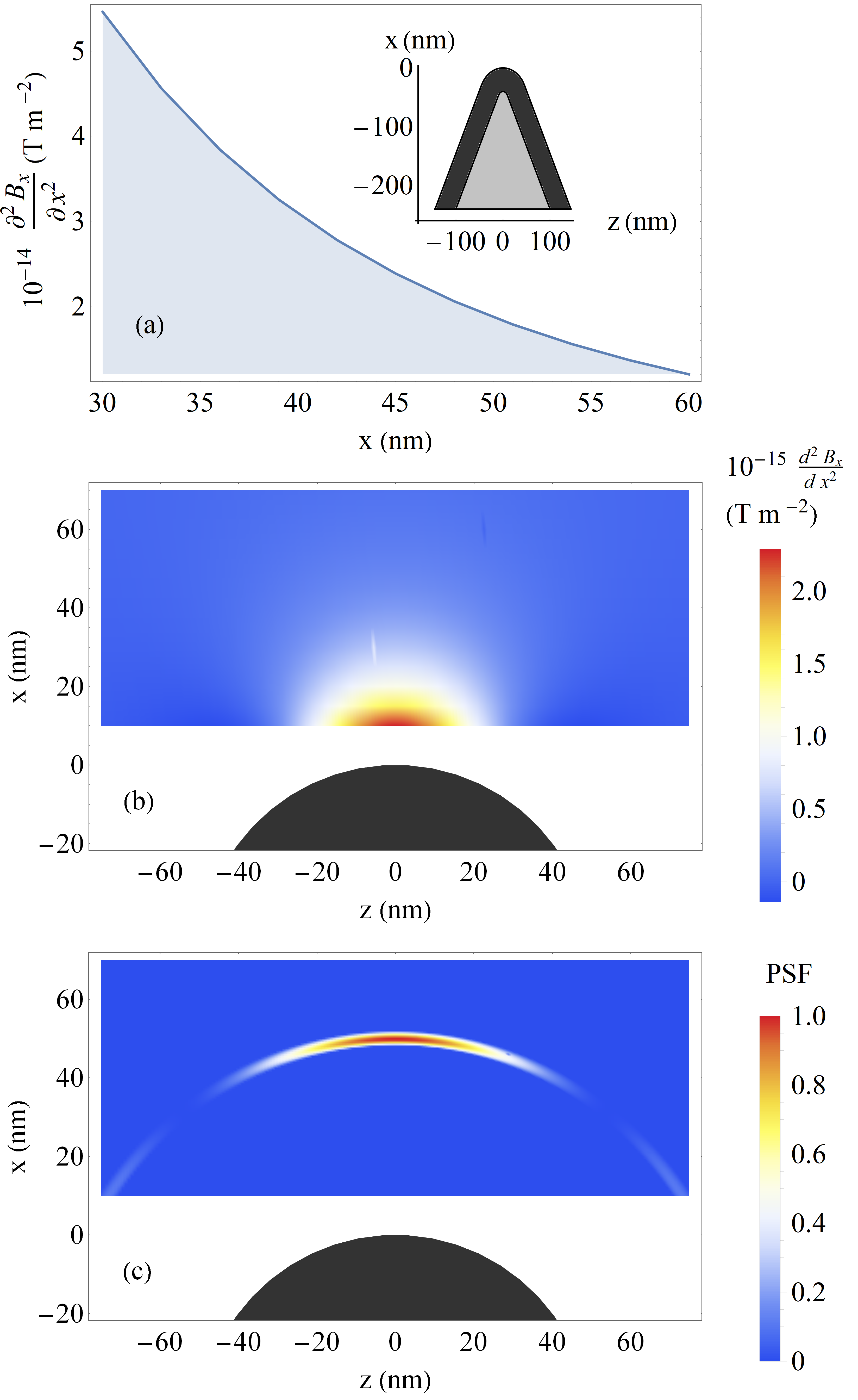}
    \caption{Magnetic field modelling results. (a) The second field gradient along the central axis of the magnetic tip; (inset) cross-section of the tip model, consisting of a non-magnetic conical base (light grey) and a layer of magnetic material (dark grey); (b) simulated second field gradient in the central plane of symmetry; (c) simulated PSF [cf. Eq.~\eqref{eq:PSF}], normalized to the value at 50\;nm above tip center. $\Delta \omega_{\rm rf} / \gamma_n = $ 10 mT was used.}
    \label{fig:fields}
\end{figure}

The magnetic field profile was estimated by modelling a hollow conical tip with a rounded top [cf. Fig.~\ref{fig:fields} (a)] with the magnetostatics package RADIA~\cite{Chubar_1998}. The tip was assumed to be magnetized to 1.83\;T parallel to the $x$-axis~\cite{Grob_2019}. A plot of the spatial profile of the second field gradient $\frac{\partial^2 B_x}{\partial x ^2}$ is shown in Fig.~\ref{fig:fields} (b). At 50\;nm above the tip center, we obtain $\frac{\partial ^2 B_x}{\partial x^2} = 2\cross10^{14}$ T\;m\textsuperscript{-2}.  

In an MRFM experiment, the spin-containing voxels constituting the sample cannot be scanned individually in real space. Instead, the frequency $\omega_{\rm rf}$ of the RF spin-flipping field is swept from $\omega_{\rm rf,0}-\Delta\omega_{\rm rf}$ to $\omega_{\rm rf,0}+\Delta\omega_{\rm rf}$. Spins whose Larmor frequency lies within these bounds are flipped, producing a signal proportional to the second field gradient, cf. Sec.~\ref{sec:model}. The signal magnitude due to a spin at position $\vb{r}$ is hence a function of space, known as the point spread function (PSF). Here, we define the PSF as
\begin{equation} \label{eq:PSF}
  \text{PSF}(\vb{r}) = \bigg[\frac{\partial^2 B_x (\vb{r})}{\partial x^2}\bigg]^2
  \bigg[1-\bigg(\frac{\gamma_n |\vb{B}(\vb{r})| - \omega_{\rm rf,0}}{\Delta\omega_{\rm rf}}\bigg)^2\bigg]
\end{equation}
for $\gamma_n |\vb{B}(\vb{r}) - B_0| \leq \Delta \omega_{\rm rf}$ and 0 otherwise, with $\gamma_n$ being the nuclear spin gyromagnetic ratio. The bracketed term is an empirical expression describing flipping fidelity, whereby spins further off the central resonant condition produce less signal~\cite{Degen_2009}.

We plot the PSF in Fig.~\ref{fig:fields} (c), using for $\omega_{\rm rf,0}$ the Larmor frequency 50\;nm above the tip center. Note that in conventional MRFM, where the transducer is a cantilever moving along the $z$-axis, the relevant gradient would be $\frac{\partial B_x}{\partial z}$, which results in PSF maxima near the edges of the tip~\cite{Degen_2009}. With our proposed method based on $\frac{\partial B_x}{\partial x}$, these maxima persist but cannot be used due to the vertical motion of the membrane. We however find an additional active area on the central axis of the magnetic tip which makes for a feasible sample position. 

\section{Spin dynamics on the moving membrane} \label{app:spins}
A conceivable drawback of our scheme is the impact of the high pump mode amplitude, as well as the thermal motion of the membrane, on the spin ensemble. Since the ensemble moves rapidly through a region with a field gradient, its lifetime may be decreased by undergoing non-adiabatic dynamics. In particular, the effect of thermal noise has previously been found important in the context of cantilever-based MRFM~\cite{Berman_2003, Mozyrsky_2003}. 

We describe the flipping in the frame rotating with the Larmor frequency about the $x$-axis, where, under the effect of an applied RF field $B_{\rm rf}(t) \cos[\omega_{\rm rf} (t)]\: \hat{\vb{e}}_z$, the effective field reads
\begin{equation}
\vb{B}_{\rm rf}(t) = \mqty(\omega_{\rm rf}(t) / \gamma_n \\0 \\ B_{\rm rf}(t) )
\end{equation}
with the spin dynamics being governed by the Bloch equation,
\begin{equation} \label{eq:bloch}
\dot{\vb{M}}(t) = \gamma_n \vb{M}(t) \cross \vb{B}_{\rm rf}(t) .
\end{equation}
Starting with $\vb{B}_{\rm rf} (t)$ parallel to $x$-axis, a spin-flip is achieved by an adiabatic sweep across the Larmor frequency. For simplicity, we take a sinusoidal RF profile,
\begin{equation}
\vb{B}_{\rm rf}(t) = B_{\rm rf}\mqty(\cos{\Delta \omega t} \\ 0 \\ \sin{\Delta \omega t})
\end{equation}
with $B_{\rm rf} = $ 5\;mT~\cite{Grob_2019} and $\Delta\omega = \omega_A - \omega_S$ = $5\cross10^4$\;s\textsuperscript{-1}. Optimizing the pulse profiles will likely provide even more stable spin inversions~\cite{Grob_2019}.
\\

\textbf{Motion of the pump mode.} When the pump mode oscillates with amplitude $X_S$ [cf. Eqs.~\eqref{eq:transform}~and~\eqref{slowflow}], the sample position is $x_1(t) =  X_S \cos{\omega_S t} / \sqrt{2}$. Such a motion is equivalent to a spurious time-dependent field
\begin{equation}
    \vb{\delta B}(t) = \frac{X_S}{\sqrt{2}} \frac{\partial B_x}{\partial x} \cos{\omega_S t}\: \hat{\vb{e}}_x \;.
\end{equation}
We solve Eq.~\eqref{eq:bloch} numerically with the field $\vb{B}_{\rm rf}(t) + \delta \vb{B}(t)$. From the field modelling in Appendix~\ref{app:fields}, we obtain $\frac{\partial B_x}{\partial x}=6\cross10^6$ T~m\textsuperscript{-1}.

\begin{figure}
    \centering
    \includegraphics[width=\columnwidth]{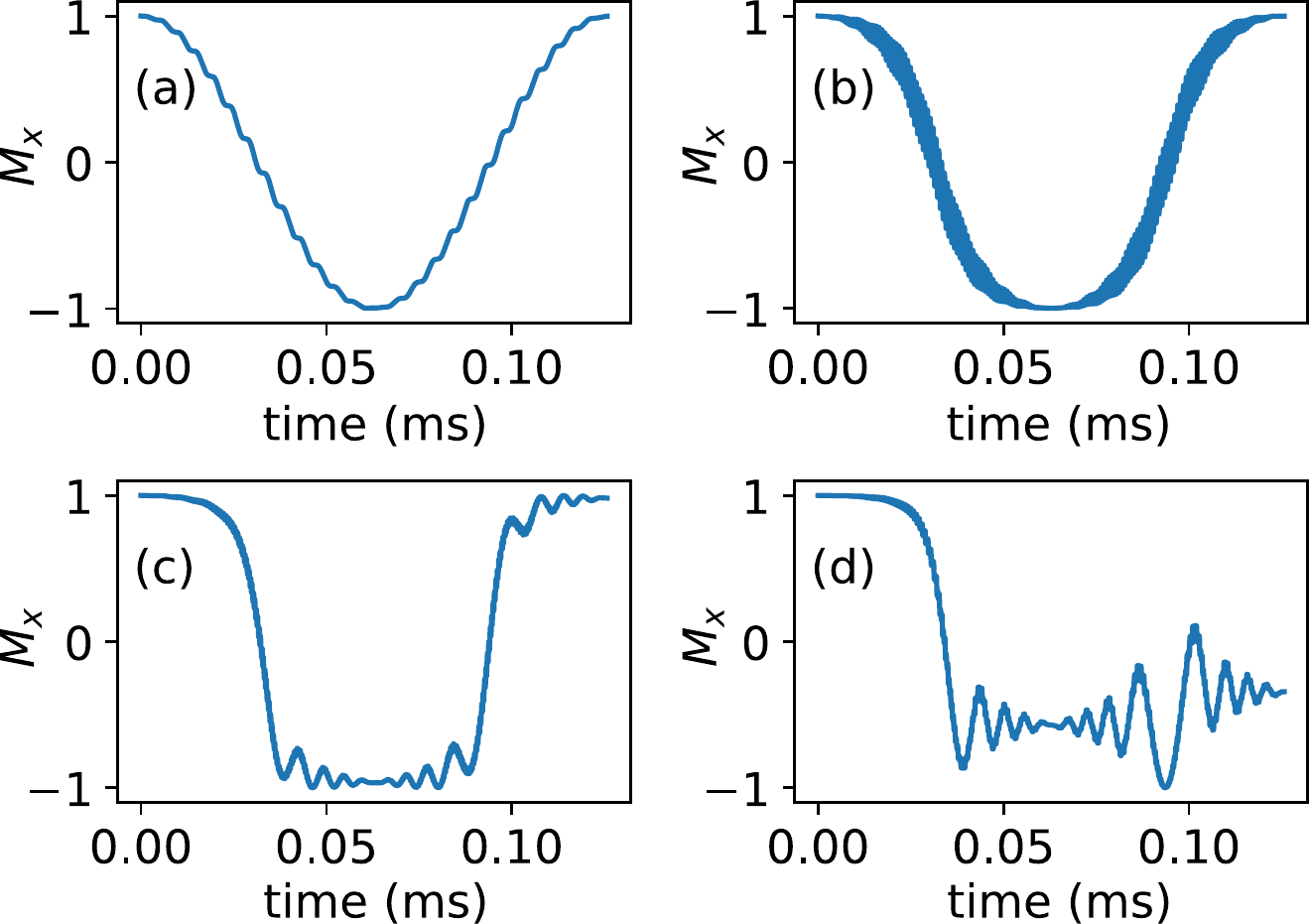}
    \caption{Magnetization component along the $x$-axis in the flipping process under increasing values of the pump amplitude. (a) $X_S$ = 0, (b) $X_S$ = 10\;nm, (c) $X_S$ = 50\;nm, (d) $X_S$ = 100\;nm.}
    \label{fig:flips}
\end{figure}

Fig.~\ref{fig:flips} shows the flipping process under increasing values of $X_S$. We observe that the spurious field induces oscillatory features in the flipping process, but only causes significant distortion at very strong ($X_S \approx$ 100\;nm) pumping.

Finally, to test the flipping fidelity, we integrated Eq.~\ref{eq:bloch} over 250 flips using $X_S=$ 10\;nm. Starting with a unit vector $\vb{M}(0)=\hat{\vb{e}}_x$, the magnetization $M_x$ at the end of each flip never dropped below 0.996. We thus conclude the flipping mechanism remains robust under strong driving of the pump mode.
\\

\textbf{Thermal noise in the membrane.} We measured the thermal displacement on one of the defect mode sites of a Si\textsubscript{3}N\textsubscript{4} membrane. At room temperature, a root-mean-square displacement of 150~pm was observed, most of which was due to the many delocalized modes of the membrane. Since cantilever-based MRFM displays high flipping fidelities at comparable displacement noise levels, and since our envisioned operational temperature (0.2 K) will further reduce thermal fluctuations, we do not expect this to be an issue with regards to spin-flipping.
\\

\section{Reference values} \label{app:vals}
All resonator parameters used in Sec.~\ref{sec:disc} are shown in Table~\ref{table:1}. The values are taken from recent experimental data~\cite{Catalini_2020}. Note that a different, non-unitary normal mode transformation is typically used in experimental literature,
\begin{equation}
\mqty(x_S \\ x_A) = \mqty(1 && 1 \\ 1 && -1) \mqty(x_1 \\ x_2)\,. 
\end{equation}
Relative to our notation, this scales the cubic nonlinearities $\alpha, \Gamma_{\rm nl}$ by a factor of $\frac{1}{2}$ and the mass $m$ by a factor of 2. This transformation is convenient for experimental use, but requires additional renormalization when dealing with external forces.

The magnetic field gradients were estimated by modelling a conical magnetic tip made of saturated NdFeB  magnet (such as is used in MFM) with the magnetostatics package RADIA~\cite{Chubar_1998}. The sample is assumed to be positioned directly above the center of the magnetic tip, where there is a relatively large area of constant $\partial^2_x B$. 

\begin{widetext}
\label{table:1}
\begin{center}
\begin{table} [h!]
\caption{Reference resonator parameters and magnetic field characteristics.}
\def\arraystretch{1.3}
\begin{tabular}{|  m{3cm} |  m{2cm} | m{8cm} |}
\hline
$m$ (ng) & 1 & resonator mass  \\ 
\hline
$\omega_0$ (s$^{-1}$)& $8.2 \cross 10^6$ & resonator natural frequency\\ 
\hline
$Q$ & $10^8$ & quality factor  \\
\hline
$\alpha$ (kg m$^{-2}$ s$^{-2}$) & $1 \cross 10^{12}$ & coefficient of Duffing nonlinearity \\
\hline
$\omega_A - \omega_S$ (s$^{-1}$) & $5 \cross 10^4$ & normal mode frequency splitting\\
\hline
Var($x_1$) (m\textsuperscript{2}) & $2.2 \cross 10^{-20}$ &  thermal displacement variance at room temperature \\

\hline
$\partial_x B$ (T m$^{-1}$) & $6 \cross 10^{6}$ & magnetic field gradient $\frac{\partial B_x}{\partial x}$\\
\hline
$\partial_x^2 B$ (T m$^{-2}$) & $2 \cross 10^{14}$ & second magnetic field gradient $\frac{\partial^2 B_x}{\partial x^2}$ \\
\hline
$\eta$ & $0.5$ & detection efficiency \\
\hline
$S_{\rm det}$ (m$^2$~s) & $10^{-31}$ & detector noise PSD \\
\hline
Var($\omega_A^2$) (s$^{-4}$) & 10$^{-1}$ & frequency variance due to intrinsic frequency noise\\
\hline
$S_\chi(\omega_A -\omega_S )$ (s$^{-1}$) & $\leq 10^{-36} $ & relative intrinsic frequency noise PSD at resonance \\
\hline
\end{tabular}
\end{table}
\end{center}
\end{widetext}

\clearpage
\bibliography{bibliography}

\end{document}